\documentclass{article}
\usepackage{graphics}
\title{\bf Brane russian doll}
\author{ Pedro F. Gonz\'{a}lez-D\'{\i}az.\\
Instituto de Matem\'{a}ticas y F\'{\i}sica Fundamental\\ Consejo Superior
de Investigaciones Cient\'{\i}ficas\\ Serrano 121, 28006 Madrid,
SPAIN\\ }
\date{May 9, 2001}
\begin{document}
\maketitle
\large
\setlength{\baselineskip}{0.9cm}

\begin{center}
{\bf Abstract}
\end{center}

It is shown that an ($n-1$)-dimensional inflating brane world
instantonically created from nothing can exist in the region
beyond the Rindler horizon of the Lorentzian spacetime
associated with another inflating brane world in $n$ dimensions
which is also instantonically created from nothing. Generalizing
this construction we obtain an unbounded from above tower of
successive brane worlds inside brane worlds, each having one
more dimension than the one which it nests and one less
dimension than the one which nests it.

\pagebreak

The large recent influx of papers on brane-world cosmology (for
a review see [1]) clearly shows the stir produced in the
physical community by the two seminal papers by Randall and
Sundrum [2,3] in 1999 on the physics of extra dimensions. These
two papers were precedented by related ideas put forwards by
Rubakov and Shaposhnikov [4], Visser [5] and Arkani-Hamed,
Dimopoulos and Dvali [6]. In the spirit of the cosmological
Garriga-Sasaki model [7], the present paper deals with the
proposal of the instantonic creation of an inflating brane world
inside the unobservable Lorentzian region of another larger
brane. This new proposal would allow us to consider a novel
mechanism for dimensional reduction and, when trivially extended
to larger (even infinite) towers of branes inside branes, our
model may accommodate a gradually induced resolution of the
hierarchy problem [2,3].

Let us start by considering a four-dimensional brane-world
embedded in an anti-de Sitter (AdS) bulk whose six-dimensional
Euclidean metric can be written in the form
\begin{equation}
ds^2=dr^2+\rho_0^2\sinh^2(r/\rho_0)\left(d\chi^2+\sin^2\chi
d\Omega_4^2\right) ,
\end{equation}
where $d\Omega_4^2$ is the metric on the unit four-sphere, and
$\rho_0=(-10/\Lambda_6)^{1/2}$, with $\Lambda_6$ the
six-dimensional cosmological constant. If we were going to
consider a compact brane-world in five-dimensional spacetime,
this could then be described by means of an instanton
constructed following e.g. the cutting-pasting procedure
introduced by Garriga and Sasaki [7], that is by excising the
spacetime region defined by the radial extra coordinate $r$ at
values larger than a given arbitrary value $r_0$, and gluing two
copies of the remaining spacetime along the five-sphere at the
given $r=r_0$. A four-brane can then be introduced on the
hypersurface at $r_0$ if we provide that brane with a tension
\begin{equation}
\sigma=\frac{3}{4\pi G_6\rho_0}\coth\left(r_0/\rho_0\right) ,
\end{equation}
so that the Israel's matching conditions [8] are satisfied. The
resulting instantonic configuration can be interpreted as a
semiclassical path for the creation of a five-dimensional
universe from nothing. Clearly this is a mathematical nice model
without direct application to the observable universe which we
live in, and exactly corresponds merely to the five-dimensional
extension of the Garriga-Sasaki instanton. Similarly, one can
cut our five-dimensional instanton in half to get a solution
that interpolates between nothing, at the south pole, and a
four-dimensional spherical brane of radius
$H^{-1}=\rho_0\sinh\left(r_0/rho_0\right)$, at the equator. This
is the five-dimensional de Sitter instanton with the inside of
the brane filled with an AdS bulk.

On the other hand, one can also consider the evolution of the
resulting brane after its creation as being described by the
analytical continuation of metric (1) such that [7]
$\chi\rightarrow iHt+\pi/2$. This gives
\begin{equation}
ds^2=dr^2+\rho_0^2\sinh^2\left(r/\rho_0\right)\left[-H^2 dt^2
+\cosh^2\left(Ht\right)d\Omega_4^2\right].
\end{equation}
On the brane at $r=r_0$, metric (3) reduces to $ds^2=-dt^2
+H^{-2}\cosh^2\left(Ht\right)d\Omega_4^2$, which represents an
inflating five-dimensional de Sitter space. It does not cover
the whole of the Lorentzian spacetime, but only the exterior
region of the Rindler horizon at $r=0$. Covering also the
interior region to complete description of the whole spacetime
requires the complexification of the coordinates $r$ and $t$:
$r\rightarrow it_c$, $Ht\rightarrow r_c -i\pi/2$, with which the
metric becomes
\begin{equation}
ds^2=-dt_c^2+\rho^2\sin^2\left(t_c/\rho_0\right)\left[dr_c^2
+\sinh^2 r_c d\Omega_4^2\right] .
\end{equation}
This is the line element that describes a six-dimensional open
FRW universe with the coordinate $r_c$ running from 0 to
$\pi\rho_0$, at which surface there is a Cauchy horizon
connecting to a new spacetime. However, as it was also noted by
Garriga and Sasaki in the five-dimensional case [7], in the
present six-dimensional framework, even though we were living in
a five-dimensional universe, the extended solution (4) would be
nothing but a mathematical idealization, describing a spacetime
which is physically unreacheable from the assumed physical
region external to the Rindler horizon that contains the brane.
Nevertheless, the usual four-dimensional instantonic picture of
the birth from nothing of an inflating de Sitter universe,
according to the no boundary condition, can also be recovered
from metric (4). In fact, starting with our six-dimensional
spacetime (1), it is always possible to construct a compact
3-brane-world instanton inside the continuation ($r\leq
r_0\rightarrow it_c$, with $t_c\leq t_{c0}$) of the above
higher-dimensional brane instanton by excising the spacetime
regions at $t_c\geq t_{c0}$ and at $r_c\geq r_{c0}$ (with
$t_{c0}$ and $r_{c0}$ being given arbitrary values of
coordinates $t_c$ and $r_c$), and gluing two copies of the
remaining spacetime along the four-sphere at $t_c=t_{c0}$,
$r_c=r_{c0}$. This can be readily seen by noting that, on the
hypersurface at $t_c =t_{c0}$, metric (4) reduces to
\begin{equation}
ds^2=dz^2
+\tilde{\rho}_0^2\sinh^2\left(z/\tilde{\rho}_0\right)d\Omega_4^2
,
\end{equation}
where $z=\tilde{\rho}_0 r_c$, with
\begin{equation}
0\leq\tilde{\rho}_0=\rho_0\sin\left(t_{c0}/\rho_0\right)\leq
\rho_0 .
\end{equation}
This is the line element of a five-dimensional Euclidean AdS
space. On this hypersurface, one can repeat the construction of
a brane-world, this time a four-dimensional world with a
three-brane, by simply exising the spacetime region $z
>z_0$, and glueing two copies of the remaining spacetime. On
$z=z_0$ a brane with tension $\sigma_5=\frac{3}{4\pi
G_6\tilde{\rho}_0}\coth\left(z_0/\tilde{\rho}_0\right)$ can
finally be introduced. The associated instanton would represent
the path for the creation of a four-universe from nothing in the
region beyond the Rindler horizon corresponding to a brane-world
with one more dimension. Together with the fact that the final
brane-world is filled with an AdS bulk, this distinguishes the
present model from the original Garriga-Sasaki scenario. Again
the Lorentzian evolution of the observable brane after creation
should be given by the analytical continuation of one of the
spherical coordinates of the unit metric $d\Omega_4^2$. Taking
this to be expressed as $d\psi^2+\sin^2\psi d\Omega_3^2$, one
can continue coordinate $\psi$ such that $\psi\rightarrow
i\tilde{H}\tilde{t}+\pi/2$, in which $\tilde{H}^{-1}=
\tilde{\rho}_0\sinh\left(z_0/\tilde{\rho}_0\right)$, to induce
the following metric
\begin{equation}
ds^2=dz^2+\tilde{\rho}_0^2\sinh^2\left(z/\tilde{\rho}_0\right)\left[-\tilde{H}^2
d\tilde{t}^2
+\cosh^2\left(\tilde{H}\tilde{t}\right)d\Omega_3^2\right].
\end{equation}
The conformal diagram of the Lorentzian spacetime corresponding
to the observable brane-world spacetime, a finite region of
which (but not all) is described by this metric, is given in
Fig.1. Such a brane world at $z=z_0$ is an inflating de Sitter
space with Hubble rate $\tilde{H}^{-1}$:
\begin{equation}
ds^2=-d\tilde{t}^2+
\tilde{H}^{-2}\cosh^2\left(\tilde{H}\tilde{t}\right)d\Omega_3^2
.
\end{equation}

\begin{figure}[h]
\begin{center}
\scalebox{.9}[.8]{\includegraphics*{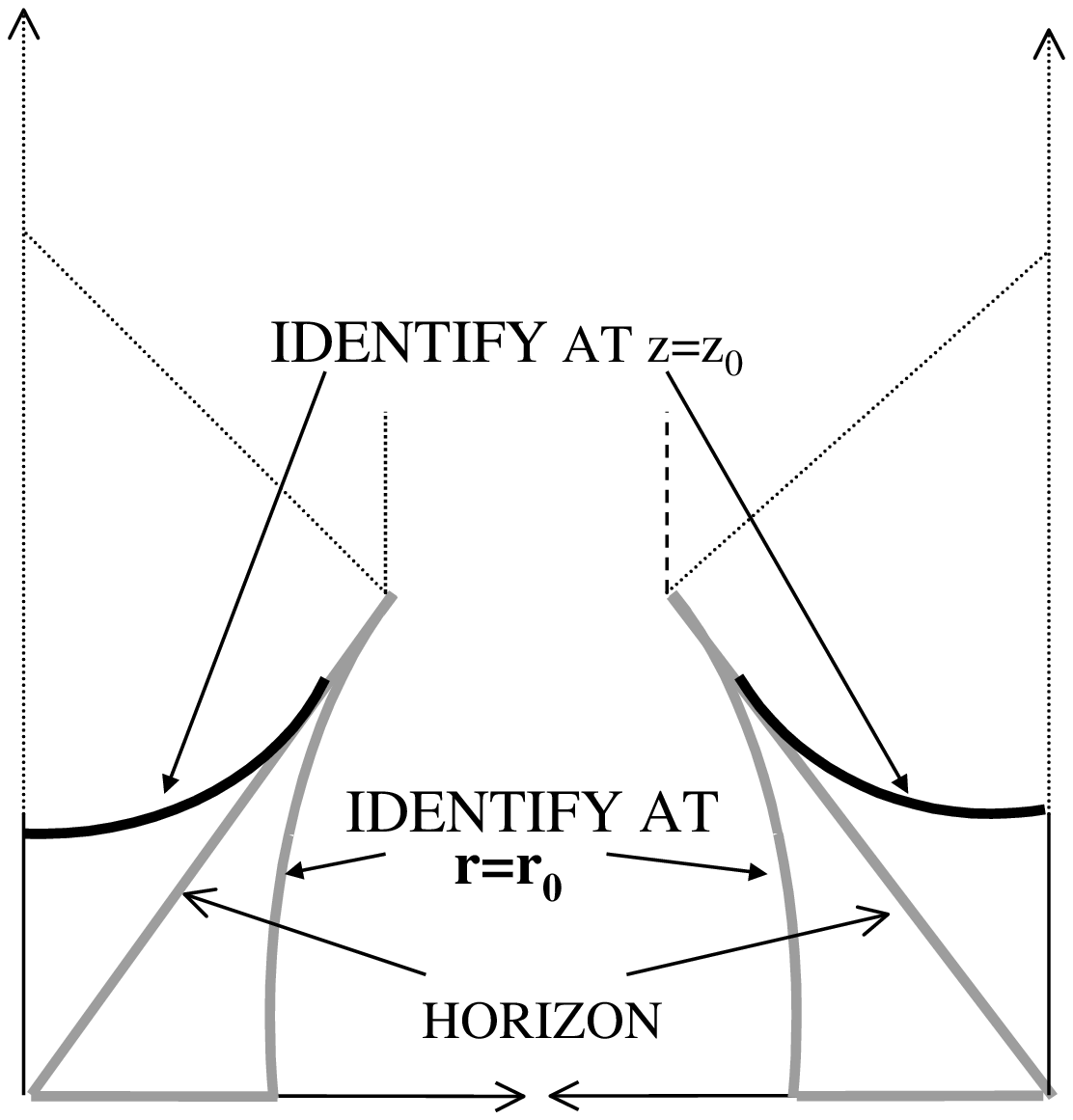}}
\end{center}
\caption{Conformal diagram of the inflating four-brane world (at
$z=z_0$) inside an also inflating five-brane world (at $r=r_0$).
Each point in the figure represents a four-sphere.}
\end{figure}

This resulting observable brane has the same bound state for
gravity and the same spectrum of Kaluza-Klein excitations as
those obtained starting with a five-dimensional Euclidean AdS
spacetime [7], with an energy gap given now by $(3/2)\tilde{H}$.
We have thus been able to construct a four-dimensional de Sitter
brane-world instanton starting with a six-dimensional AdS
instanton, which is localized in the region beyond the event
horizon of the five-dimensional Lorentzian brane. What matters
now is to compute the action of the four-dimensional brane-world
instanton within the realm of the five-brane instanton. This can
be accomplished by summing the Euclidean actions corresponding
to the brane-worlds in the six- and five-dimensional AdS spaces,
starting with the nonlinear generalization of the zero mode for
the respective brane [2,3]. Thus, for the Euclidean
six-dimensional AdS space (with topology ${\rm R}^1\times{\rm
S}^5$) we shall use the metrical ansatz
\begin{equation}
ds^2=g_{ab}dx^a dx^b =dr^2+a(r)^2\gamma_{\mu\nu}dx^{\mu}dx^{\nu}
,
\end{equation}
with $\gamma_{\mu\nu}$ the five-dimensional metric and
$a(r)=\rho_0\sinh\left(r/\rho_0\right)$. In order to avoid the
singular character of the one-brane solution and prepare the
system to accommodate a cosmological scenario, we shall use a
configuration made of by two conccentrical branes which would
cosmologically evolve independently of each other. To the brane
at $r=r_0$, we thus add a new brane with a negative tension
given by reversing sign of Eq. (2) at some radius $r_1 <r_0$,
excising the spacetime region $r>r_1$ and identifying the edges
of the two copies of AdS space bulk along the five-sphere at
$r=r_1$ [7]. Then, expressing the surface term $\int d^5
x\sqrt{h}{\rm Tr}K$ in terms of the brane's tension, $\sigma$,
i.e. ${\rm Tr}K=32\pi G_6\rho_0 \sigma/3$, we have for the
Euclidean action of the system formed by two conccentrical
branes with tensions $\sigma_i$ ($i=0,1$), respectively located
at $r_i$ and separated by a region of AdS bulk described by the
generalized metric (9)
\begin{equation}
S_E^{(6)}= \frac{1}{16\pi G_6}\int d^6 xa^5\left(2\Lambda_6
-R\right)\sqrt{\gamma^{(5)}}+\frac{4}{3}\sum_i\sigma_i a^5
V_5^{(\gamma)} ,
\end{equation}
where $V_5^{(\gamma)}=\int d^5 x\sqrt{\gamma^{(5)}}$ is the
five-volume. Introducing an energy-momentum tensor
\begin{equation}
T_{ab}=-\frac{g_{ab}\Lambda_6}{8\pi G_6} -\sum_i a^2 \sigma_i
\gamma_{ab}^{(5)}\delta(r-r_i) ,
\end{equation}
in which $\gamma_{ab}^{(5)}=0$ provided that $a$ or $b$ equals
$r$, eliminating the scalar curvature $R$ from the trace of the
Einstein equation, $R=3\Lambda_6 +20\pi G_6 \sum_i \sigma_i$,
and expliciting the six-dimensional cosmological constant as
$\Lambda_6=-10/\rho_0^2$, we finally obtain, after performing
the integral,
\begin{equation}
S_E^{(6)}= -\sum_{i=0}^{1}(-1)^i
\cosh\left(r_i/\rho_0\right)\left[\frac{1}{\sinh^2\left(r_i/\rho_0\right)}
+\frac{15}{32}\sinh^2\left(r_i/\rho_0\right)
-\frac{1}{2}\coth^2\left(r_i/\rho_0\right)\right]S_E^{(6)i} ,
\end{equation}
where
\begin{equation}
S_E^{(6)i}= -\frac{2V_5^{(\gamma)}}{3\pi G_N H_i^2} ,
\end{equation}
with $G_N=G_6/\rho_0^2$ the Newtonian gravitational constant,
and $H_i=a_i^{-1}=
1/\left(\rho_0\sinh\left(r_i/\rho_0\right)\right)$. The
calculation of the Euclidean action for the generalized
five-dimensional AdS instanton follows a similar pattern, to
finally produce:
\begin{equation}
S_E^{(5)}= -\sum_{j=0}^{1}(-1)^j
\left[\frac{z}{\tilde{\rho}_0\sinh^2\left(z/\tilde{\rho}_0\right)}
-\coth\left(z/\tilde{\rho}_0\right)
+\frac{4}{15}\cosh\left(z/\tilde{\rho}_0\right)
\sinh\left(z/\tilde{\rho}_0\right)\right]S_E^{(5)j} ,
\end{equation}
where
\begin{equation}
S_E^{(5)j}=-\frac{5V_4^{(\gamma)}}{8\pi G_N \tilde{\rho}_0
H_j^2} ,
\end{equation}
with $V_4^{(\gamma)}=\int d^4 x\sqrt{\gamma^{(4)}}$ the
four-volume, $\gamma_{\mu\nu}$ the four-dimensional metric
entering the ansatz for the generalized line element $ds^2=
dz^2+a^2(z)\gamma_{\mu\nu}dx^{\mu}dx^{\nu}$,
$G_N=G_6/\tilde{\rho}_0^2$, and
$H_j=1/[\tilde{\rho}_0\sinh\left(z/\tilde{\rho}_0\right)]$. The
probability for the creation of a five-brane instanton
containing a four-brane instanton itself created in the interior
region of the Lorentzian continuation of the first instanton
will be then proportional to $\exp(S_E)$, with $S_E=S_E^{(6)}+
S_E^{(5)}$. It can be checked that such a probability will
closely depend on the combined valued for the fixed coordinates
$r$ and $z$ at the brane locations.

The above instantonic construction can be readily extended to
encompass any $n$-dimensional starting Euclidean metric,
$ds^2=dr^2+\rho_0^2\sinh^2(r/\rho_0)\left(d\chi^2+\sin^2\chi
d\Omega_{n-2}^2\right)$, which, if we follow such a construction
for every $p$-dimensional Lorentzian interior region (with
$n\leq p\leq 5$) beyond the event horizon of the successive
branes, will lead to a russiandoll-like tower of inflating brane
worlds. In such a construct, the innest brane should be chosen
to be the lowest dimensional and the outest brane would be
$(n-1)$-dimensional, and each brane would be fully unobservable
for the rest of branes in the whole tower. Inflating
brane-worlds could thus exist inside the unobservable region of
higher dimensional brane-worlds which would also be inflating.
Our own universe could in this way be just one (the lowest
dimensional or not) among a large (perhaps infinite) tower of
universes with different number of physical dimensions, which
would be geometrically unobservable for us. The creation of each
of these universes should be from nothing only in the sense of
ausence of the spacetime of its own observable region, but other
even large regions outside the given universe's horizon would be
allowed to exist at its origin. In any event, while the brane
scenario presented in this paper offers the opportunity: of
establishing a new and perhaps advantageous mechanism for
dimensional reduction or increase, to allow a gradual solution
of the hierarchy problem, and shed new light at the concept of
creation from nothing in cosmology, the application of the
holographic idea [9] to the creation of de Sitter universes by
the procedure described in this work leads to the encoding on
the cosmological horizon of only a finite number of degrees of
freedom [10], in contradiction [11] with string and M theories.
One would then adhere to a pure higher dimensional
general-relativity interpretation of the brane worlds rather
than adcribing a string- or M-theory origin for such worlds.

\vspace{.8cm}

\noindent{\bf Acknowledgements}
The author thanks C.L. Sig\"uenza and A. Zhuk for many useful
discussions. This work was supported by DGICYT under Research
Project No. PB97-1218.

\pagebreak

\noindent\section*{References}

\begin{description}
\item [1] C. Barcel\'{o} and M. Visser, JHEP 0010 (2000) 019.
\item [2] L. Randall and R. Sundrum, Phys. Rev. Lett. 83 (1999)
3370.
\item [3] L. Randall and R. Sundrum, Phys. Rev. Lett. 83 (1999)
4690.
\item [4] V. Rubakov and M. Shaposhnikov, Phys. Lett. B125
(1983) 136.
\item [5] M. Visser, Phys. Lett. B159 (1985) 22.
\item [6] N. Arkani-Hamed, S. Dimipoulos and G. Dvali, Phys.
Lett. B429 (1998) 263.
\item [7] J. Garriga and M. Sasaki, Phys. rev. D62 (2000)
043523.
\item [8] W. Israel, N. Cimento B44 (1966) 1; B48 (1966) 463.
\item [9] R. Bousso, JHEP 0011 (2000) 038.
\item [10] R. Bousso, {\it Bekenstein bounds in de Sitter and
flat space}, hep-th/0012052.
\item [11] W. Fischler, A. Kashani-Poor, R. McNees and S. Paban,
{\it The Acceleration of the Universe, a Challenge for String
Theory}, hep-th/0104181.

\end{description}

\end{document}